\documentclass[conference]{IEEEtran}
\usepackage{times}
\usepackage{helvet}
\usepackage{courier}

\usepackage[pdftex]{graphicx}
\usepackage[pdftex]{color}
\usepackage{url}
\usepackage{comment}
\usepackage{caption}
\DeclareCaptionType{copyrightbox}
\usepackage{subcaption}

\def\etal{et~al.\,}
\def\ie{\emph{i.e.}\,}

\def\naive{na\"{\i}ve }

\begin{document}
\title{Online Social Media in the Syria Conflict: \\Encompassing the Extremes and the In-Betweens}


\author{
Derek O'Callaghan$^1$, Nico Prucha$^2$, Derek Greene$^1$, Maura Conway$^3$, 
\\ 
Joe Carthy$^1$, P\'{a}draig Cunningham$^1$ \\ \\
$^1$School of Computer Science \& Informatics,
University College Dublin, Email: \{name.surname\}@ucd.ie\\
$^2$Department of Near Eastern Studies,
University of Vienna, Email: nico.prucha@univie.ac.at\\ 
$^3$School of Law \& Government,
Dublin City University, Email: maura.conway@dcu.ie
}

\maketitle
\begin{abstract}
The Syria conflict has been described as the most socially mediated in history, with online social media playing a particularly important
role. At the same time, the ever-changing landscape of the conflict leads 
to difficulties in applying 
analytical 
approaches taken by
other studies of online political activism.
Therefore, 
in this paper, 
we use an approach that does not require strong prior assumptions or the proposal of an advance hypothesis
to analyze Twitter and YouTube activity of a range of protagonists to the conflict, in an attempt to reveal
additional insights into the relationships between them. By means of a network representation that combines multiple data views, we uncover communities
of accounts falling into four categories that broadly reflect the situation on the ground in Syria. A detailed analysis of selected
communities within the anti-regime categories is provided, focusing on their central actors, preferred online platforms, and 
activity surrounding ``real world'' events. 
Our findings indicate that social media activity in Syria is considerably more convoluted than reported in many other studies of online
political activism, suggesting that alternative analytical approaches can play an important role in this type of scenario.
\end{abstract}

\section{Introduction}

The conflict in Syria, which has been ongoing since March 2011, has been characterized by extensive use of online social media platforms
by all sides involved \cite{APSyriaPlaysOut2013}. 
For example, YouTube is being used by a variety of alleged Syrian sources to document and highlight events as they occur,  
where over a million videos have been uploaded since January
2012 according to official figures, which in turn have received hundreds of millions of views \cite{WSJSyriaRealTime2013}.
Consequently, this behaviour suggests that any related studies should consider multiple online platforms, and in this paper we analyze the activity of a range
of protagonists to the conflict on two of the most popular platforms used, namely Twitter and YouTube. Our primary interest is to understand
the extent to which this online activity may reflect the known situation on the ground, where additional insights are revealed into the
relationships between the corresponding groups and factions.

In this paper, we consider the Syria conflict within the context of studies of online political activism, where attention is often focused
upon relatively static (often mainstream) groupings about which a considerable level of prior knowledge is available; for example, divisions
between liberal and conservative US communities, or mainstream European political parties
\cite{Adamic:2005:PBU:1134271.1134277,TumasjanPredictingElections2010,Jungherr:2012:WPP:2206736.2206740}. The static nature of these groups
can provide a certain advantage in terms of driving data sampling strategy such as using a pre-defined set of Twitter hashtags, or defining a
specific research question such as the prediction of electoral results using online activity. In the case of the Syria conflict, a \naive application of these standard approaches might be to initially assume the existence of a pro/anti-regime dichotomy. However, the constantly shifting
landscape of the conflict suggests that an alternative approach is worthy of consideration \cite{FPSyriaMultiPolar2013}. 
In an effort to avoid making strong prior assumptions or proposing an advance hypothesis, 
we perform an analysis of the complex Syrian situation as follows:

\begin{enumerate}
    \item a directed collection of Twitter data associated with the Syria conflict that leverages existing authoritative
    sources.
	\item a set of groups is generated from the collected data, where we achieve this by means of community detection within a
	corresponding network representation.
	\item a conceptual high-level categorization of the detected communities is proposed.
	\item a detailed interpretation of selected communities is provided, where we are particularly interested in demonstrating
	the complex nature of the conflict, in contrast to other online political environments.
\end{enumerate}

\begin{figure*}[!t]
    \centering
    \includegraphics[width=0.72\linewidth]{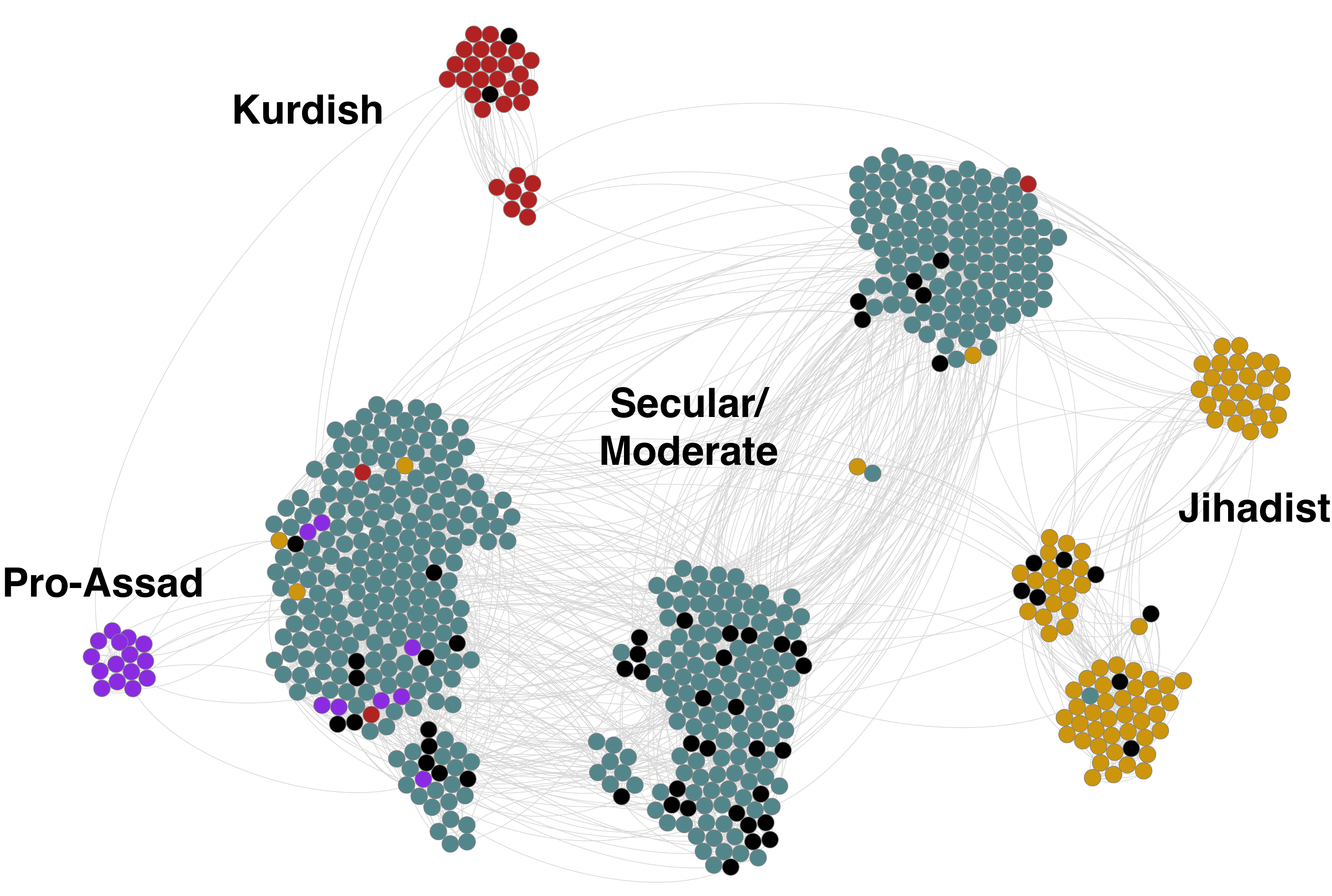}
	\caption{Syrian account network (652 nodes, 3,260 edges). Four major categories; Jihadist (gold, right), Kurdish (red, top), 	Pro-Assad
	(purple, left), and Secular/Moderate opposition (blue, center). Black nodes are members of multiple communities. Visualization was
	performed with the OpenOrd layout in Gephi.}
    \label{fig:macrocomms}
\end{figure*}

To this end, we generate a unified network representation that combines multiple data views \cite{Greene:2013:PUG:2464464.2464471}, within
which communities of Twitter accounts are detected. Although other studies of Syrian Twitter activity have analyzed large-scale data sets \cite{USIPSyriaSociallyMediated2014},
here we employ a smaller curated set of Syrian accounts originating from a variety of Twitter lists, where we consider these accounts as
having been deemed authoritative or interesting in some way by the list curators. A similar focus on authoritative sources has been employed
by other entities that monitor social media activity associated with the conflict, for example, Eliot Higgins, the maintainer of
the Brown Moses blog\footnote{\url{http://brown-moses.blogspot.com}}, who monitors hundreds of YouTube channels daily
with a view to verifying the content of videos uploaded by a variety of Syrian sources \cite{GigaOMBrownMoses2013}.

We find that the communities within this Syrian network can be placed into four major
categories (see Fig.~\ref{fig:macrocomms}). A detailed analysis of three selected communities within the anti-regime categories is provided later in this paper, focusing on their central actors and preferred online platforms. To illustrate the contrast with the polarization analyzed 
in certain studies of mainstream political activism \cite{Adamic:2005:PBU:1134271.1134277,ConoverICWSM112847}, the three communities selected consist of two polar opposites, jihadist and secular revolutionary, with the third community
considerably moderate in comparison. The analysis process includes the generation of rankings of the preferred YouTube channels for
each community, where these channels and corresponding Freebase topics assigned by YouTube are used to assist interpretation
while also providing a certain level of validation\footnote{\url{http://www.freebase.com/}}.
We also consider online activity surrounding ``real world''
events, such as  YouTube video responses to the Ghouta chemical weapon attack on 21 August 2013 \cite{BBCGhouta2013}. The insights revealed in this
study confirm that alternative analytical approaches can play a key role in studies of online activity
where prior knowledge may be scarce or unreliable.

\section{Analyzing Online Political Activism}

In this paper, we consider online activity associated with the Syria conflict within the context of other studies of online political
activism that have focused upon relatively static, often mainstream groupings about which a considerable level of prior knowledge is
available. This includes situations featuring a polarization effect, or others where multiple groupings are in existence.
For example, the study of US liberal and conservative blogs by Adamic and Glance \cite{Adamic:2005:PBU:1134271.1134277} found clear
separation between both communities, with noticeable behavioral differences in terms of network density based on links between blogs, blog
content itself, and interaction with mainstream media. They did not focus on ``other'' blogs, such as those of a libertarian, independent or
moderate nature (and found few references to these from the liberal and conservative blogs), but suggested that they could be considered in
future analysis.
Progressive and conservative polarization on Twitter was investigated by Conover \etal, where hashtags were used to gather data
leading to two network representations based on Twitter retweets and mentions \cite{ConoverICWSM112847}. By specifically requesting the detection of
exactly two communities, polarization was clearly observable in the retweet network. This was not the case with analogous two-community
detection within the corresponding mentions network, where the authors suggested that this feature may foster cross-ideological interactions
of some nature. In both cases, increasing the number of target communities beyond two revealed smaller politically heterogeneous communities
rather than those of a more fine-grained ideological structure.

Mustafaraj \etal analyzed the \emph{vocal minority} (prolific tweeters) and \emph{silent majority} (accounts that tweeted only once)
within US Democrat and Republican Twitter supporters, gathering data by searching for tweets containing the names of two Massachusetts
senate candidates \cite{DBLP:conf/socialcom/MustafarajFWM11}. They also found similar polarized retweet communities in the vocal minority, while at the same time, the
activity of both of these communities was consistently different to the silent majority at the opposite end of the spectrum.
The machine learning framework proposed by Pennacchiotti and Popescu for the classification of Twitter accounts was evaluated using three
gold standard data sets, including one associated with political affiliation that was generated from lists of users who classified
themselves as either Democrat or Republican in the Twitter directories WeFollow and Twellow \cite{Pennacchiotti:2011:DRS:2020408.2020477}.
Similar political affiliation on Twitter was studied by Wong \etal, where they proposed a method to quantify US political
leaning that focused on tweets and retweets \cite{WongICWSM136105}. Here, data was collected using keywords associated with the Democrat and
Republican candidates in the 2012 presidential election. They suggested that their approach compared favourably to follower or retweet graph-based analysis due to the
difficulty in interpreting the results of the latter.
Hoang \etal also focused on Twitter political affiliation around the time of
the 2012 presidential election, using a binary classifier (Democrat and Republican) that did not consider ``neutral'' (other) affiliation
\cite{HoangPoliticsMicroblogs2013}. 

Gruzd and Roy found evidence of both polarization and non-polarization in their analysis of tweets posted prior to the 2011 Canadian federal
election \cite{GruzdPolarizationCanadian2014}. In this case, tweets were selected using a single election hashtag, and accounts were
manually classified based on affiliation with one of the established political parties, with supporters of non-mainstream parties being excluded. In a similar approach to that of
Conover~\etal, Weber \etal studied the polarization between Egyptian secular and Islamist accounts on Twitter \cite{WeberSecularIslamistAsonam2013}, with a set of seed accounts that had been manually labelled as such being used to drive data collection. Among their findings, they observed a certain level of polarization in the retweet network, albeit at a lower level than that observed by Conover \etal in the corresponding progressive and conservative network. Regarding online Syria activity, Morstatter \etal used a set of Syrian hashtags in their analysis of Twitter data sampling
approaches \cite{MorstatterICWSM136071}. 
More recently, the study of Syrian Twitter behaviour by Lynch \etal included an analysis of macro-level
activity over time \cite{USIPSyriaSociallyMediated2014}, using tweets from the firehose Twitter API containing the word ``Syria'',
regardless of the source or account location. Unlike the situation in the US, polarization was not observed within the retweet network,
where a number of high-level communities both within Syria and elsewhere were identified.

Prediction of election results has also been the focus of studies of political social media activity. For example, Tumasjan
\etal examined the power of Twitter as a predictor of the 2009 German elections, using tweets mentioning the six main political parties
currently in parliament at the time, along with a selection of prominent politicians \cite{TumasjanPredictingElections2010}. They found that
the ranking of parties generated using the volume of associated tweets was identical to the actual ranking in the election results, with
relative volume mirroring the corresponding share of votes received by each party. However, other commentators have voiced concerns about over-estimating Twitter's predictive power; in
particular, Jungherr \etal suggested that if the Pirate Party (not in parliament at the time) had been included in the analysis of Tumasjan
\etal, they would instead have won the 2009 election based on relative frequency of mentions within tweets
\cite{Jungherr:2012:WPP:2206736.2206740}. They discuss the requirement for well-grounded rules when collecting data.
Similarly, Metaxas \etal found that using Twitter to predict two US elections in 2010 generated results that were only slightly better than chance
\cite{MetaxasNotPredictElections2011}.

The studies discussed above have largely focused on relatively static and often mainstream political situations about which a certain
level of prior knowledge is available, for example, activity related to supporters of US Democrats and Republicans. This can act as a
driver in the collection of large data sets and corresponding analysis. In other situations, such as that of the Syria conflict, current and reliable knowledge may be lacking.
Consequently, we propose that smaller-scale studies of related online activity
using methods that do not necessitate the development of (possibly incorrect) assumptions, as is the case in this paper, are appropriate.

\section{Data}

For the purpose of this analysis, we collected a variety of data for a set of Twitter accounts
associated with the Syria conflict. To find relevant accounts, we made use of the Twitter \textit{user lists} feature, which is often employed by journalists and other informed parties
to curate sets of accounts deemed to be authoritative on a particular subject. This method has previously been shown to be successful in
exploiting the wisdom of the Twitter crowds in the search for topic
experts \cite{Greene:2012:ACN:2365934.2365941,Ghosh:2012:CCS:2348283.2348361}.
Two approaches were taken: 1) we identified lists of Syrian accounts by known journalists, and 2) we identified lists containing one or more
of a small number of official accounts for recognized high-profile entities known to be active on the ground, such as the Free Syrian Army
(FSA) and the Islamic State of Iraq and al-Sham [greater Syria] (ISIS). The latter consisted of lists curated by both Syrian and non-Syrian
journalists, academics, and Syrian entities directly involved in the conflict.

Having aggregated the accounts from these lists into a single set, we then proceeded to manually analyze each profile and filter immediately
identifiable accounts such as non-Syrian media outlets and journalists, academics and think-tanks, non-governmental organizations (NGOs),
and charities. The intention was to specifically retain Syrian accounts that claimed to be directly involved in events in Syria, or Syrians
that were commentating on events from abroad. Accounts whose nature could not be ascertained, or were not publicly accessible, were also
excluded. Although there was a potential for bias to be introduced with this process, we felt that it was preferable to identify accounts
to be excluded rather than specifying which accounts were to be included, where we trusted the list curators' judgement in the case of the
latter. At the same time, it should be mentioned that any samples of online social media activity will inevitably contain a certain level of
selection bias, regardless of the sampling strategy employed, for example, the use of pre-defined Twitter hashtag sets or other methods
\cite{TufekciICWSM2014}.

Using a set of 17 Twitter lists yielded a
total of 911 unique accounts, from which a final set of 652 accounts remained following the filtering process. Twitter data including followers, friends, tweets and list memberships were retrieved for each of the selected accounts during October and November 2013, as limited by the Twitter API restrictions effective at the time. 1,760,883 tweets were retrieved in total, with 21\% and 75\% of these being posted in 2012 and 2013 respectively.
Follower relations between the final set of accounts yielded 30,808 links, while 175,969 mentions and 27,768 retweets were also identified
pertaining to these accounts. In addition, Twitter list memberships were available for 647 accounts, associated with 22,080 distinct lists.
All valid YouTube URLs were also extracted from the tweet content, and profile data was retrieved for the corresponding video and channel
(account) identifiers using the YouTube Data API, resulting in a set of 14,629 unique channels that were directly/indirectly tweeted by 619
unique accounts. As YouTube automatically annotates uploaded videos with Freebase topics where possible\footnote{See
\url{http://youtu.be/wf_77z1H-vQ} for an explanation of this process.}, we also retrieved all available topic assignments for videos
uploaded by these channels.

\section{Methodology}

To detect communities, we created a network representation using the method proposed by Greene and
Cunningham \cite{Greene:2013:PUG:2464464.2464471}, which was previously applied to aggregate multiple network-based and content-based \textit{views} or relations from Twitter into a single \emph{unified graph}. Given a data set of $l$ different views, the unified graph method consists of two distinct phases:
\begin{enumerate}
\item \emph{Neighbor identification:} For each view, we compute the similarities between every account $u_{i}$ and all other accounts in
that view, using an appropriate similarity measure (in this paper we use cosine similarity, due to the sparsity of the data matrices). We use
these similarity values to rank the other accounts relative to $u_{i}$. After repeating this process for all views, the $l$ resulting
rankings are merged together using SVD rank aggregation \cite{wu10recsys}. From the aggregated ranking, we select the $k$ highest ranked
accounts as the \emph{neighbor set} of $u_{i}$. This is repeated for all accounts in the data set.
\item \emph{Unified graph construction:} We then build a global representation of the data set by constructing the corresponding asymmetric
$k$-nearest-neighbor graph.  That is, a directed unweighted graph where an edge exists from account $u_i$ to $u_j$, if $u_{j}$ is contained in
the neighbor set of $u_{i}$. This results in a sparse unified graph encoding the connectivity information from all the original views,
covering all accounts that were present in one or more of those views.
\end{enumerate}

In this work, we employ the unified graph approach to combine views of Twitter accounts related to the Syria conflict, based on information derived from both Twitter and YouTube. This is a combination of explicit network relations, and implicit network representations that are derived from tweet content and Twitter lists. Specifically, we use seven Twitter-based views constructed from the data described previously to generate the required neighbor rankings.
\begin{enumerate}
\item \emph{Follows:} From the unweighted directed follower graph, construct binary account profile vectors based on accounts whom they
follow (\ie out-going links).
\item \emph{Followed-by:} From the unweighted directed follower graph, construct binary account profile vectors based on the accounts that
follow them (\ie incoming links). A pair of accounts are deemed to be similar if they are frequently ``co-followed'' by the same accounts.
\item \emph{Mentions:} From the weighted directed mention graph, construct account profile vectors based on the accounts whom they mention.
\item \emph{Mentioned-by:} From the weighted directed mention graph, construct account profile vectors based on the accounts that mention
them.
A pair of accounts are deemed to be similar if they are frequently ``co-mentioned'' by the same accounts.
\item \emph{Retweets:} From the weighted directed retweet graph, construct account profile vectors based on the accounts whom they retweet.
\item \emph{Retweeted-by:} From the weighted directed retweet graph, construct account profile vectors based on the accounts that retweet
them.
Accounts are deemed to be similar if they are frequently ``co-retweeted'' by the same accounts.
\item \emph{Co-listed:} Based on Twitter list memberships, construct an unweighted bipartite graph, such that an edge between a list
and an account indicates that the list contains the specified account.  A pair of accounts are deemed to be similar if they are frequently
assigned to the same lists.
\end{enumerate}
Due to the high volume of user-generated video content associated with the Syria conflict \cite{WSJSyriaRealTime2013}, we have also constructed an
additional \emph{YouTube channel view}. This view reflects the extent to which pairs of Twitter accounts post tweets containing links to the same YouTube videos or channels. In this respect, the view is analogous to the co-listed Twitter view. Note that we aggregate the YouTube information at the channel level due to the well-documented existence of authoritative channels \cite{GigaOMBrownMoses2013}. 

For all of the above views, pairwise similarity values between account vectors were computed using cosine similarity, which in turn allowed
us to identify each account's neighbors. From these neighbor sets, we then produced a single unified network where each account has at
most $k=5$ neighbors, as used previously in \cite{Greene:2012:ACN:2365934.2365941}. This yielded a network consisting of 652 nodes and 3,260 edges. We then applied the OSLOM community detection algorithm to this network \cite{lancichinetti11oslom}. OSLOM finds overlapping communities, which is appropriate here, as previous work has shown that partitioning algorithms can fail to accurately return community structure in social media networks \cite{Reid2011}. Due to the effect of the resolution parameter $P$ on the number and size of communities found by OSLOM, where lower values result in larger numbers of smaller communities, multiple runs were executed for values of $P$ in [0.1, 0.9]. In each case we specified that all nodes were to be assigned to communities. A manual inspection of the communities found in each run was
performed to discover the value of $P$ leading to the smallest number of larger communities. This led to the selection of $P=0.8$ (higher values led to communities being overly merged). Details on these communities are provided in the next section.

\begin{figure*}[!t]
    \centering
    \includegraphics[width=0.94\linewidth]{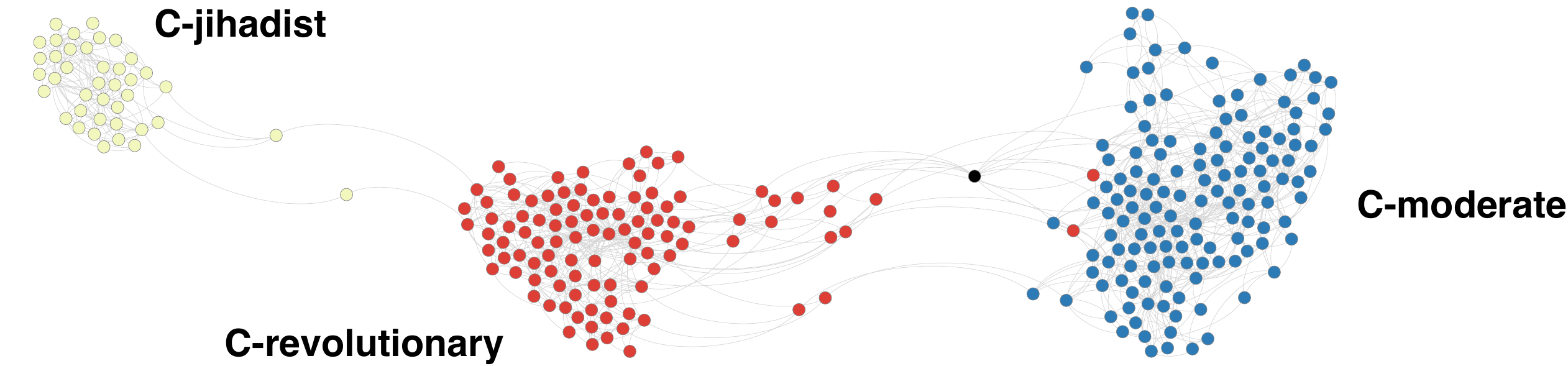}
    \vskip 0.5em
	\caption{The induced subgraph for the three communities selected for detailed analysis: \emph{C-jihadist} (40 nodes, yellow, left), \emph{C-revolutionary} (105
	nodes, red, center), and \emph{C-moderate} (137 nodes, blue, right). Only one node (black) in this subgraph was assigned to multiple
	communities (\emph{C-revolutionary} and \emph{C-moderate}). This account presents itself as the board of all local coordination committees
	in Syria, which comprise activists from a range of religious beliefs and ethnic heritage that report on events as they occur.
	Visualization was performed with the Force Atlas layout
	(repulsion strength = 50.0) in Gephi.}
    \label{fig:subgraph}
\end{figure*}

To assist in the interpretation of the resulting OSLOM communities, for each community we generated rankings of the YouTube channels that were
present in the corresponding member tweets (channels were tweeted directly or indirectly via specific video URLs). A ``profile document''
was generated for each account node, consisting of channel identifiers extracted from their tweets. These were then represented by log-based
TF-IDF vectors normalized to unit length. For each community, the subset of vectors for the member accounts was used to calculate a mean
vector $D$, with the channel ranking consisting of the top 25 channel identifiers in $D$.
A second set of community rankings was generated, based on the automatic Freebase topic annotation of videos by YouTube. A ``topic
document'' was created for each unique channel that featured in $\geq1$ community channel ranking, consisting of an aggregation of the
English-language labels for all topics assigned to their respective uploaded videos. As before, a ranking was generated from the top topics
of the corresponding $D$ vector.

\section{Syria Community Categories}

We now focus on the communities identified by the complete process described previously.
In all, 16 communities were found, where inspection of inter-community linkage determined
the presence of four major categories. The separation between these categories can be seen in the visualization produced with the OpenOrd
layout in Gephi in Fig.~\ref{fig:macrocomms} \cite{bastian09gephi,martin2011openord}.
The four categories are:

\begin{enumerate}
	\item \textit{Jihadist}: three communities that include accounts with a stated commitment to violent 
	jihadism, including accounts associated with or supportive of 
	the al-Nusra Front (Jabhat al-Nusra) and the Islamic State of Iraq and al-Sham (ISIS).

	\item \textit{Kurdish}: two communities, with one community containing a variety of accounts that include political parties and local media
	outlets, while the other appears to be mostly associated with youth movements and organizations.

	\item \textit{Pro-Assad}: one community, where accounts are primarily used to declare support for the current Syrian government and armed
	forces. Notable entities include the Syrian Electronic Army (SEA), a collection of hackers that target opposition groups and western
	websites.

	\item \textit{Secular/Moderate opposition}: ten communities, containing accounts for various opposition entities such as the National
	Coalition for Syrian Revolutionary and Opposition Forces (Syrian National Coalition) and the Free Syrian Army (FSA). These communities
	constitute a dense core of the network, with many common channels and topics found within the corresponding community rankings.
\end{enumerate}

The largely Arabic language accounts were categorized manually on the basis of a lingual and visual analysis of their content, in addition
to analysis of individual egocentric networks. There are distinct and immediately apparent differences recognizable by subject matter
experts and Arabic speakers between the keywords, political and theological concepts, and symbols used by violent jihadists versus those
employed by pro-Assad accounts, for example \cite{SyrianConflictReconciliation2014}.

A ranking of the top 25 YouTube channels was produced for each community, with 295 unique channels found across all 16 rankings, from a
total of 14,629 that were extracted from tweets. For each of these 295 channels, a random sample of up to 1,000 uploaded videos was
generated, since each YouTube API call to retrieve videos uploaded by a channel returns a maximum of 1,000 results. A topic-based vector was then created using all Freebase topics assigned to this video sample. The mean percentage of sampled videos having annotated topics was 59\% ($\sigma = 27\%$) per channel. In total, 185 unique topics were found
across the resulting community topic rankings.

\begin{table*}[!t]
        \caption{Statistics reported by Twitter for the accounts in the three communities selected for interpretation.}        
        \begin{subtable}{1.0\textwidth}
        \centering
		\begin{tabular}{| l | r | r | r | r | r | r |}
		\hline 
		\textbf{Community} & \textbf{Total} & \textbf{Min} & \textbf{Max} & \textbf{Mean} & \textbf{Median} & \textbf{Standard deviation}
		\\
		\hline \hline 
		C-revolutionary \hspace{1em} & 5,092,982 & 22 & 438,285 & 48,504 & 15,889 & 77,284 \\ \hline
		C-jihadist & 230,712 & 33 & 23,705 & 5,767 & 2,899 & 6,339 \\ \hline
		C-moderate & 2,469,995 & 19 & 162,989 & 18,029 & 10,292 & 24,252 \\ \hline
		\end{tabular}
		\caption{Tweet statistics.}
		\label{tab:communitytweets}
        \end{subtable}

        \begin{subtable}{1.0\textwidth}
        \centering
		\begin{tabular}{| l | r | r | r | r | r | r |}
		\hline 
		\textbf{Community} & \textbf{Total} & \textbf{Min} & \textbf{Max} & \textbf{Mean} & \textbf{Median} & \textbf{Standard deviation}
		\\
		\hline \hline 
		C-revolutionary \hspace{1em} & 5,302,075 & 58 & 1,505,272 & 50,495 & 10,140 & 153,416 \\ \hline
		C-jihadist & 489,976 & 247 & 79,089 & 12,249 & 5,658 & 17,879 \\ \hline
		C-moderate & 774,917 & 7 & 136,297 & 5,656 & 2,159 & 12,832 \\ \hline
		\end{tabular}
		\caption{Follower statistics.}
		\label{tab:communityfollowers}
        \end{subtable}
\label{tab:communitystats}
\end{table*}

\section{Interpretation of Communities}

In this section, we offer a detailed analysis of a selection of communities, based on observation of the associated Twitter profiles,
tweets, retweets, related accounts, YouTube channel profiles, and uploaded video content. To illustrate the complexity of the situation on
the ground, while also demonstrating how a \naive assumption of a pro/anti-regime polarization is unsuitable, we have selected three diverse
communities within the anti-regime categories, consisting of two communities from the Jihadist and Secular/Moderate categories which may
themselves be considered as polar opposites of each other, with the third community of a far more moderate nature. These are hereafter
referred to as \emph{C-jihadist}, \emph{C-revolutionary} and \emph{C-moderate} respectively. A visualization of the induced subgraph for
these communities can be found in Fig.~\ref{fig:subgraph}, where
the edges between C-revolutionary and C-moderate can be explained by the presence of many secular accounts.
Table~\ref{tab:communitystats} contains the distribution of tweets and followers reported by Twitter for the three communities. Note
that, due to the Twitter API restrictions effective at the time, only a subset of total tweets reported were used in the current analysis.
Table~\ref{tab:topics} provides an initial insight into all three in terms of the highest-ranked topics assigned to their preferred YouTube
channels. While the C-revolutionary and C-moderate topics mainly refer to Syrian cities or provinces, with the Free Syrian Army featuring
strongly for the former, the C-jihadist topics reflect the community's agenda with references to The Nusra Front and ISIS, the two most
important Jihadist groups active in Syria. The common Al-Qusayr topic refers to the Syrian-Lebanese border
town that is of strategic value to both sides.

C-jihadist is the smallest of the three communities, containing just 40 accounts. The majority of these are violent jihadist in their
orientation, with a particular affinity evidenced for ISIS ideology. Language-wise this community is heavily Arabic: 95\% of all its accounts are
Arabic-only. The `black banner' and other iconography associated with violent jihadism can be observed, including images of Osama bin Laden.
Photos tweeted by these accounts include many of weaponry and attacks, also close-ups of `martyred' fighters and a small number of
individuals holding up severed human heads. One of the accounts with highest follower in-degree claims to be an official ISIS account, but
is not.
It is however maintained by a staunch ISIS supporter, has accumulated nearly 87,000 followers, and provides a link on the account's
marquee to a Q\&A page on jihadist ideology on ask.fm. The majority of account holders are also active on jihadi forums and reflect the
interlinked nature of this form of propaganda on a wide range of platforms, including Facebook, YouTube, Instagram, etc. Twitter is also an
important gateway to YouTube for this community. The highest ranked YouTube channel was established in summer 2013, has over 6,500
subscribers and 1.7 million total views for 1,000$+$ videos, most of which reflect the day-to-day situation on the ground in the mainly
Sunni sectors of Aleppo under attack. This channel is interesting too because, despite being heavily linked-to by members of this jihadi
community and containing some pro-jihadi content, it also posts video openly critical of ISIS.

\begin{table}[!t]
\caption{Top ten Freebase topics assigned to the preferred YouTube channels of the three selected communities.}
\begin{center}
\begin{tabular}{| l  | p{3.0cm} | p{1.8cm} |}
\hline 
 { \textbf{C-revolutionary}} & { \textbf{C-jihadist}} & { \textbf{C-moderate}}
\\
\hline \hline 
{ {Free Syrian Army}} & { {The Nusra Front}} & {
{Syria}} \\
\hline 
{ {Syria}} & { {Aleppo}} & {
{Syrian civil war}}  \\
\hline 
{ {Homs}} & { {Islamic State of Iraq and Syria}} & {
{Aleppo}}  \\ 
\hline 
{ {Aleppo}} & { {Earth}} & {
{Damascus}}  \\
\hline 
{ {Damascus}} & { {Anasheed} }& {
{Free Syrian Army}}  \\
\hline 
{ {Idlib}} & { {Ali}} & {
{Homs}}  \\
\hline
{ {Al-Qusayr}} & { {Buraidah}} & {
{Bashar al-Assad}}  \\
\hline 
{ {Rif Dimashq Governorate}} & { {Iraq}} & {
{Al-Rastan}} \\ 
\hline 
{ {Daraa}} & { {Al-Safira}} & {
{Al-Qusayr}}  \\
\hline 
{ {Al-Rastan}} & { {Al-Qusayr}} & {
{Kafr Batna}}  \\
\hline
\end{tabular}
\end{center}
\label{tab:topics}
\end{table}

C-revolutionary, containing 105 nodes, is composed largely of Free Syrian Army (FSA) supporters and sympathizers and contains the most
prolific tweeters of the three communities. Like C-jihadist, it too is heavily Arabic language-centric, with some 83\% of accounts containing
Arabic-only tweets. It also shares the dissemination of still images of attacks and their aftermaths, lynchings, and the bodiless heads of alleged non-Syrian
suicide bombers. One of the most prominent accounts is an official account of the Syrian Revolutionary Forum, which is explicitly anti-ISIS
and paints the current conflict as ``from the sky [we face] the barrels of Bashar [al-Assad] and the car bombs of al-Baghdadi [ISIS leader]
on the ground.'' The account, which has nearly 73,000 followers, also documents the aftermath of attacks and supplies photographs of
unidentified bodies along with requests for help in identifying them. Another prominent account, which has some 30,000 followers, is
maintained by an individual committed to religious tolerance. The user maintaining this account - almost certainly a male - underlines that
the ``cross'' (i.e. Christianity) is part of his country and cites verses of the Qur'an to underline the plurality of faithful people, which
he describes as decreed by God. He also retweets conservative Sunni content however that is deployed within the jihadi Twittersphere,
unfortunately reflecting real-life grievances and sectarian violence in Syria (and Iraq). The highest ranked YouTube channel linked-to from
within this community was that of a self-described anti-regime ``news network''. 
However, the fact that this channel is no longer accessible demonstrates the potential connection between social
media activity and ``real world'' consequences.
In this case, it appears that the project's individual media activists/citizen journalists became targets for the regime's intelligence
services in their efforts to shut down this - at the time - rare professional window into the violence unleashed against unarmed civilians.

\begin{figure}[!t]
	\begin{center}
        \begin{subfigure}[b]{0.49\textwidth}
                \centering
				\hskip -0.5em
                \includegraphics[width=1.00\textwidth]{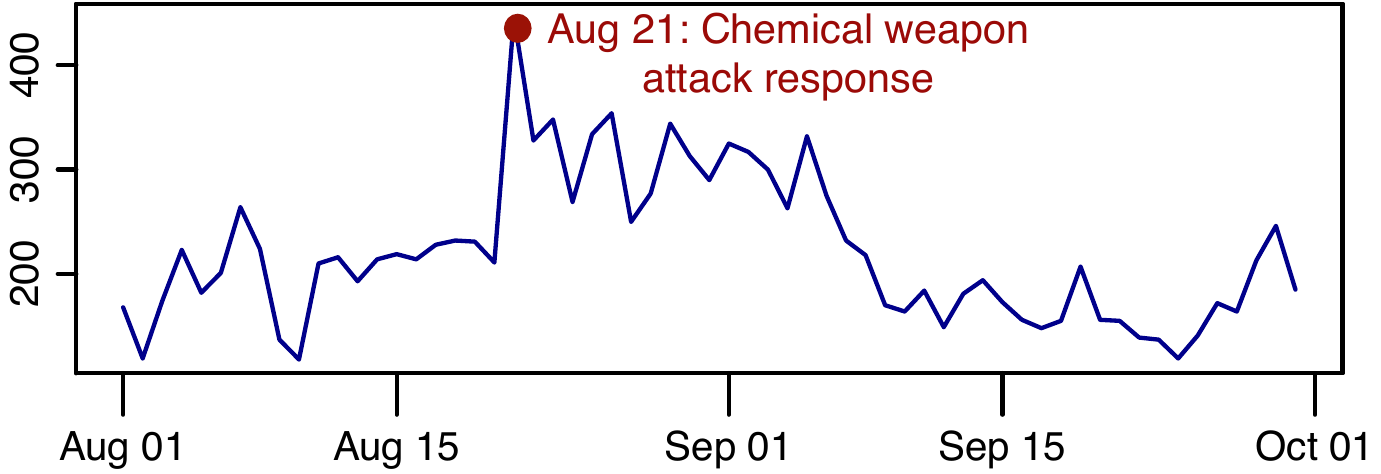}
                \caption{C-moderate}
                \label{fig:community15uploads}
        \end{subfigure}
		\vskip 2.0em
        \begin{subfigure}[b]{0.49\textwidth}
                \centering
                \includegraphics[width=1.00\textwidth]{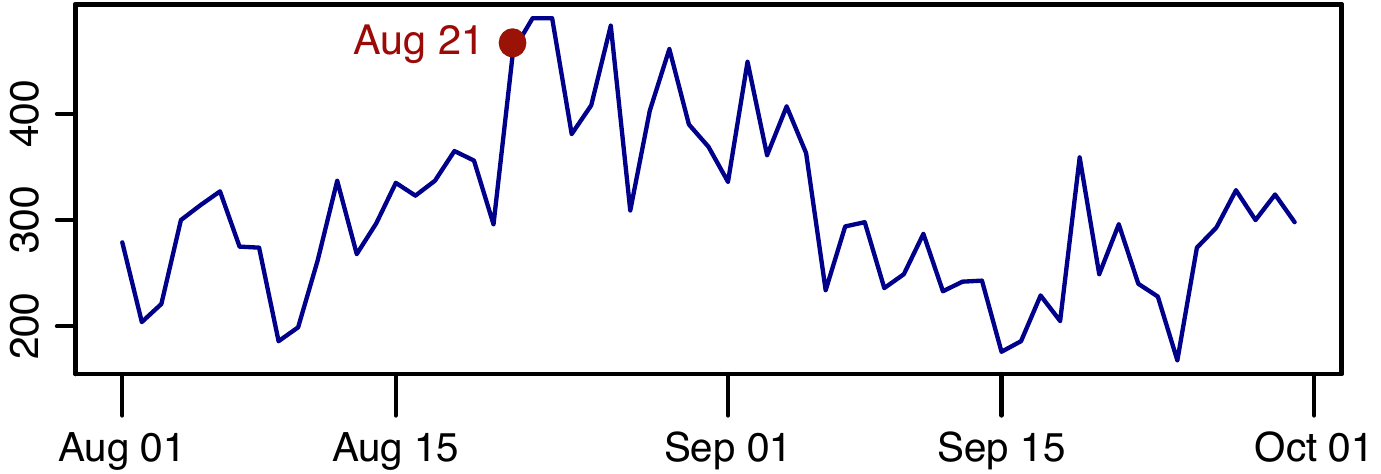}
                \caption{C-revolutionary}
                \label{fig:community2uploads}
        \end{subfigure}
		\vskip 2.0em
        \begin{subfigure}[b]{0.49\textwidth}
                \centering
                \includegraphics[width=1.00\textwidth]{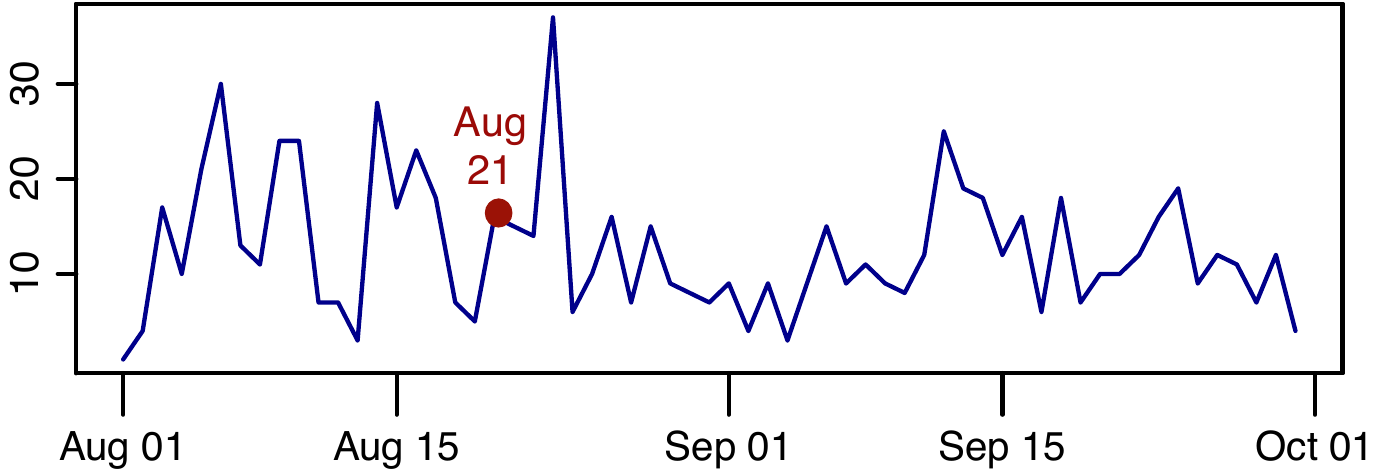}
                \caption{C-jihadist}
                \label{fig:community8uploads}
        \end{subfigure}
	\end{center}
	\caption{Video uploads by the preferred YouTube channels of the three selected communities during Aug-Sept 2013 (69 unique
	channels).}
	\label{fig:uploadactivity}
\end{figure}

C-moderate is the largest of the three communities, containing 137 accounts, with the majority maintained by a moderate and largely
secular portion of the Syrian opposition and their supporters whose tweets document the war and its daily reality on the ground in Syria. Calls for
reconciliation amongst Syrians and/or for so-called `foreign fighters' to quit Syria can also be observed. It differs from C-revolutionary
and C-jihadist in a number of ways. First, most of the accounts contain English-only tweets (60\%) or are bilingual Arabic-English (27\%).
Second, many appear to be members of the Syrian diaspora, identifying themselves as presently located not in Syria but elsewhere in the Middle East
(e.g. Jordan, Lebanon, Turkey, UAE), or in the US or Europe. Third, while C-revolutionary and C-jihadist tweet some images of children, including
those of injured or dead children and babies, these are the majority of images disseminated by C-moderate. Photos of fighters or weaponry
are rarely observed whereas photos of the Kafranbel\footnote{Kafranbel is a town in northwestern Syria whose residents have creatively responded to
the Syria conflict by creating and displaying often humorous signs that express outrage at the world's indifference to the conflict and that
oppose both the Assad regime and ISIS. Many of the signs are in English.} banners routinely appear; images of women are also much more
prominent here. The latter may probably be explained by our observation that users identifying as female are much more prominent in this
community, in comparison to the others. For example, just one user, a student and ``revolutionist'' with 12,000 followers, explicitly
identifying as female appears in C-revolutionary, while no users explicitly identifying as female are found in C-jihadist. All of the above
points may be illustrated with reference to a central user (8,000$+$ followers) who describes herself as a female from Damascus based in Europe, who
tweets in English and posts images of Syrian children and the Kafranbel banners, but never pictures of fighters. In general, the accounts
here link to major Western media sites rather than YouTube. Having said that, YouTube links are routinely supplied also, with the highest
ranked channel having been established early in the conflict (i.e. summer 2011). The channel presently has just shy of 4,000 subscribers,
including prominent media outlets such as Al-Jazeera English, and has accumulated over 3.3 million total views.

A major disparity becomes apparent when a comparison is undertaken between the volume of uploads for the top 25 YouTube channels linked-to
by C-jihadist versus that of both C-revolutionary and C-moderate as these relate to the ``real world''. There is a
particularly striking difference between the attention, in terms of both reaction time and volume of video uploaded, paid by both communities to the Ghouta
chemical weapon attack on 21 August 2013. Fig.~\ref{fig:uploadactivity} illustrates that it took the space of some two days for the 
C-jihadist preferred YouTube channels to respond to the attack whereas the response by the YouTube channels preferred by C-revolutionary and
C-moderate was immediate. The differences in volume of video uploaded by the channels are also worth commenting upon: there was a peak of
some 37 videos uploaded on 23 August by the C-jihadist channels, which tapered off again almost immediately; the corresponding 
C-revolutionary and C-moderate channels responded with the upload of over 400 videos on the day of the attack with increased volumes for
some 15 days thereafter.
Looking specifically at the period 21-29 August, we found that Freebase topics were assigned to 43\%, 45\% and 37\% of videos uploaded by
the C-jihadist, C-revolutionary and C-moderate channels respectively. The top ten topics (ranked by TF-IDF as before) for C-revolutionary
and C-moderate topics appear extremely relevant and include ``Ghouta'', ``Chemical weapon'', and ``Chemical warfare'', while similar topics
were not observed for C-jihadist (Table~\ref{tab:topicsaug2013}).

\begin{table}[!t]
\begin{center}
\caption{Top ten Freebase topics assigned to the preferred YouTube channels of the three selected
communities during the period August 21-29 2013 (the Ghouta chemical
weapon attack took place on 21 August).}
\label{tab:topicsaug2013}
\begin{tabular}{| p{1.8cm} | p{2.7cm} | p{2.9cm} |}
\hline 
 { \textbf{C-moderate}} & { \textbf{C-revolutionary}} & \textbf{C-jihadist}
\\
\hline \hline 
{ {Syria}} & { {Syria}} & {Aleppo} \\
\hline 
{ {Syrian civil war}} & { {Free Syrian Army}} & {Anasheed} \\
\hline 
{ {Damascus}} & { {Levant} } & {The Nusra Front} \\
\hline 
{ {Ghouta}} & { {Ghouta}} & {Eid prayers} \\
\hline 
{ {Chemical weapon}} & { {News}} & {Al-Bab} \\
\hline 
{ {Jobar}} & { {Homs}} & {Mosque} \\
\hline
{ {Aleppo}} & { {Aleppo}} & {Homs} \\
\hline 
{ {Bashar al-Assad}} & { {Rif Dimashq Governorate}} & {Dayr Hafir} \\
\hline 
{ {Chemical warfare}} & { {Darayya}} & {Aleppo International Airport} \\
\hline 
{ {Al-Rastan}} & { {Al-Rastan}} & {Salaheddine district} \\
\hline
\end{tabular}
\end{center}
\end{table}

\section{Conclusions and Future Work}

The Syria conflict has been described as ``the most socially mediated in history'' \cite{USIPSyriaSociallyMediated2014}. As multiple
platforms are often employed by protagonists to the conflict and their supporters, we have analyzed activity on Twitter and YouTube, where 
an approach that does not require a prior hypothesis 
has found communities of accounts that fall into four identifiable
categories.
These categories broadly reflect the complex situation on the ground in Syria \cite{FPSyriaMultiPolar2013}, contrasting with the relatively static nature of mainstream groupings that were the focus of other studies of online political activism.
A detailed analysis of selected anti-regime communities has been provided, where rankings of preferred YouTube channels and associated
Freebase topics were used to assist interpretation. Resources such as the latter can help to overcome some of the multilingual obstacles 
encountered in this context, while also addressing the interpretation issues that can exist with graph-based analysis
\cite{WongICWSM136105}.

We emphasize once more the difficulties involved in analyzing certain online political situations, due to a scarcity in the prior
knowledge that is commonly available to mainstream political analysis.
Here we have focused on a set of actors that have been deemed authoritative, as this approach has been taken elsewhere in the analysis of
online Syrian activity \cite{GigaOMBrownMoses2013}. The meaningful results found with close reading of tweet, video and other content
demonstrate the value of ``small data'', where valuable research insights can be found at any level \cite{doi:10.1080/1369118X.2012.678878},
while also complementing results found with analysis of larger data sets associated with the conflict
\cite{USIPSyriaSociallyMediated2014}.
In future work, we would like to continue this approach for further analysis of these groups, with a view
to monitoring the flux in group structure and ideology. 

\section{Acknowledgments}
This research was supported by 2CENTRE, the EU funded Cybercrime Centres of Excellence Network and Science Foundation Ireland (SFI) under
Grant Number SFI/12/RC/2289.

\bibliography{20140813_OCallaghanOSMSyriaConflict}
\bibliographystyle{IEEEtran}

\end{document}